\documentclass[9pt, conference]{IEEEtran}
\IEEEoverridecommandlockouts

\usepackage{pgfplots}
\usepackage{pgfplotstable}
\definecolor{bg}{HTML}{282828}    
\definecolor{fg}{HTML}{ebdbb2}    
\definecolor{red}{HTML}{cc241d}   
\definecolor{red}{HTML}{fb4934}   
\definecolor{green}{HTML}{98971a} 
\definecolor{yellow}{HTML}{d79921} 
\definecolor{blue}{HTML}{3d83b3}  
\definecolor{blue}{HTML}{5cb2d6}  
\definecolor{blue}{HTML}{fabd2f}  
\definecolor{blue}{HTML}{60a1e1}  
\definecolor{blue}{HTML}{9ae160}  
\definecolor{purple}{HTML}{b16286} 
\definecolor{aqua}{HTML}{689d6a}  
\definecolor{gray}{HTML}{a89984}  
\definecolor{orange}{HTML}{d65d0e}  

\tikzset{
  box/.style = {rectangle, draw, minimum width=2.5cm, minimum height=1.2cm, align=center},
  arrow/.style = {thick, -{Latex[length=2mm, width=2mm]}},
  label/.style = {font=\small\bfseries, above=2pt}
}

\usetikzlibrary{positioning}
\usetikzlibrary{external}
\usetikzlibrary{arrows.meta}
\usetikzlibrary{shapes.geometric}
\usetikzlibrary{calc}
\usetikzlibrary{patterns}
\usetikzlibrary{pgfplots.statistics}
\usetikzlibrary{positioning, arrows.meta, calc}
\usepackage{amsmath}
\usepackage{cite}
\usepackage{amsmath,amssymb,amsfonts}
\usepackage{algorithmic}
\usepackage{graphicx}
\usepackage{textcomp}
\usepackage{xcolor}
\usepackage{subcaption}
\usepackage{float}
\usepackage{subfloat}
\usepackage{comment}
\usepackage{multirow}
\usepackage{siunitx}
\def\BibTeX{{\rm B\kern-.05em{\sc i\kern-.025em b}\kern-.08em
    T\kern-.1667em\lower.7ex\hbox{E}\kern-.125emX}}

\begin{document}

\title{Anti-Aliasing Snapshot HDR Imaging Using Non-Regular Sensing}

\author{\IEEEauthorblockN{Teresa St\"urzenhof\"acker, Moritz Klimm, J\"urgen Seiler, Andr\'e Kaup}
\IEEEauthorblockA{Chair of Multimedia Communications and Signal Processing, \\
Friedrich-Alexander-Universit\"at Erlangen-N\"urnberg(FAU)\\
\{teresa.stuerzenhofaecker, moritz.klimm, juergen.seiler, andre.kaup\}@fau.de\\
}}
\maketitle

\begin{abstract}
Snapshot HDR imaging is essential to capture the full dynamic range of a scene in a single exposure, making it essential for video and dynamic environments where motion prevents the use of multi-exposure techniques or complex hardware set-ups. This work presents a snapshot HDR imaging sensor that is based on spatially varying apertures, implemented by combining two differently sized prototype pixels. The different light integration areas physically extend the dynamic range towards the lower end, compared to a standard high resolution sensor. 
A non-regular pixel arrangement is suggested, to mitigate aliasing and overcome a loss in spatial resolution that is associated with increased light integration area of the larger prototype pixel.
Subsequent reconstruction in the Fourier domain, where natural images can be sparsely represented allows to recover the image with high detail. 
The image acquisition approach with the proposed non-regular HDR sensor is simulated and analysed with special emphasis on the spatial resolution.
 The results suggest the snapshot HDR sensor layout to be an effective way to acquire images with high dynamic range and free from aliasing artefacts.
\end{abstract}

\begin{IEEEkeywords}
HDR, Non-Regular Sampling, Image Sensor.
\end{IEEEkeywords}

\section{Introduction}
\label{sec:Intro}
High dynamic range (HDR) imaging is essential for capturing scenes with significant luminance variations. Even though modern high-end cameras have  improved their dynamic range over the past few years, it is still not on a par with the human visual system \cite{humanVision}. A straight forward approach to increase the dynamic range relies on merging multiple exposures of the same scene which works very well for static scenes but fails to capture motion \cite{muliexposure}. For moving objects, beam splitters can be used to capture synchronised exposures with multiple sensors \cite{Beamsplitter}. However, this requires a more expensive hardware set-up and higher complexity. Alternatively, Dual-Conversion Gain (DCG) applies two different conversion gains to the same photodiode and thereby allows to capture both bright and dark areas \cite{DCG}. Yet, this snapshot approach does not change the number of collected photons and therefore does not extend the physical dynamic range \cite{DCG}. 
Another snapshot approach uses neutral density (ND) filters on individual pixels to physically extend the dynamic range by controlling the number of collected photons \cite{NDFilters}. While effective for capturing highlights, this method cannot improve dark scene acquisition, as modern cameras already approach photometric limits \cite{Non-regularOptical}.
The only way a sensor can reduce the impact of noise in darker scenes is by using pixels with larger light integration areas. Fujifilm introduced the SuperCCD SR sensor which is based on hexagonal pixels of different sizes \cite{SuperCCDSensor}. 
The large prototype pixel captures dark areas accurately, while the small pixel collects fewer photons and is thus better suited for bright scene regions.
While the SuperCCD SR sensor increases the dynamic range by 400\% it has never dominated the camera market due to major drawbacks regarding the spatial resolution, accompanied by increased manufacturing complexity due to the hexagonal pixel shapes \cite{HDRCapture}.
The loss in spatial resolution is a result of the larger pixel size, which in turn decreases the sampling distance and introduces aliasing. 
Our work is improving on this by proposing a snapshot HDR imaging sensor layout that benefits from differently sized pixels while mitigating aliasing and preserving high-resolution detail, by introducing non-regular pixel arrangements.
Consequently, the proposed sensor reduces information loss during image acquisition which usually origins from to exceeded exposures and aliasing. Thereby high image signal fidelity is provided for any further classical or ML-based image processing steps.
 The novelty of this work is posed by a HDR snapshot sensor design benefiting from
\begin{itemize}
	\item two differently sized prototype pixels that extend the sensors dynamic range physically by 300\%
	\item exploiting a non-regular arrangement of the prototype pixels to allow for an aliasing-free image acquisition.
\end{itemize}
The paper is organised as follows: first an overview on the image acquisition pipeline is given in Section \ref{sec:ImageAcquisition}. 
In Section \ref{sec:HDR}, the proposed HDR sensor is presented with its ability to increase the dynamic range and avoid aliasing artefacts. 
Finally, the simulations and results are provided in Section \ref	{sec:results}, followed by a summary in Section \ref{sec:conclusion}.

\section{Image Acquisition model}
\label{sec:ImageAcquisition}

To capture an image, the scene radiance $\Phi$ has to be converted into digital pixel values through physical and electronic integrations and processing \cite{HDRBook}.
Radiance from a scene, $\Phi_{\text{scene}}$, is focused onto the sensor by the lens, which integrates incoming light over a solid angle $\Omega$ defined by the aperture.
The shutter controls the exposure time, integrating irradiance over time.
A measure of the energy at each pixel can be obtained by a spatial integration 
over the pixel area.
In the final step, the image signal is amplified, quantized, digitized, and mapped to pixel values using a non-linear camera response function (CRF) \cite{HDRBook}.
The acquisition can be simplified as,
\begin{equation}
\begin{aligned}
I(x_0, y_0) \propto
\iint_{\mathcal{R}} \int_{\tau_\text{exp}} \int_{\Omega}
& \Phi_{\text{scene}}(x, y, t, \Omega) \, \text{d}\Omega \, \text{d}t \, \text{d}x \, \text{d}y \\
 + & \ N_\text{dynamic}(x, y) \,  
\end{aligned}
\end{equation}

where $\mathcal{R}$ defines the region of a pixel and $\tau_\text{exp}$ denotes the time of exposure. In reality, multiple noise sources, combined in $N_\text{dynamic}(x, y)$ degrade the image quality during the acquisition process. The most quality degrading noise source is usually shot noise which arises from the statistical variation in the number of photons hitting the sensor \cite{photometricLimit}. The shot noise behaviour can be modelled by a poisson distribution which is especially dominant in low-light situations and short exposures when only few photons arrive at the sensor. Other noise sources like dark current, read-out noise and thermal noise are approximated by adding gaussian noise to the acquisition model.
Further, the sensor saturation clips the measured intensity $I(x_0, y_0)$ according to,
\begin{equation}
	I_\text{saturated}(x_0, y_0) = \text{min}\left ( I_\text{max}, I(x_0, y_0) \right ).
\end{equation} 
The maximum value $I_\text{max}$ and the amount of present noise limit the dynamic range of the acquired image \cite{HDRCapture}. 

\section{Proposed High Dynamic Range Sensor Layout}
\label{sec:HDR}
\begin{figure}[t!]
    \centering
    \setlength{\tabcolsep}{0.3em}
    \begin{tabular}{cc}
        \resizebox{0.13\textwidth}{!}{\begin{tikzpicture}
    \pgfmathsetseed{72}

    \def\cells{4} 
    \def\size{4} 
    \def\smallsize{1} 

    \foreach \x in {0,1,...,\numexpr\cells-1} {
        \foreach \y in {0,1,...,\numexpr\cells-1} {
            \pgfmathsetmacro{\randx}{int(random(0,1))}
            \pgfmathsetmacro{\randy}{int(random(0,1))}
            \pgfmathsetmacro{\pattern}{ifthenelse(\randx==\randy, "north west lines", "north east lines")}

            \draw[black, thick, preaction={fill, red}, pattern=\pattern, pattern color=red!40] (\x,\y) rectangle ++(1,1);
            \draw[black, thick, preaction={fill, red!90}] (\x,\y) rectangle ++(1,1);
            \draw[black, thick, preaction={fill, blue}, pattern=\pattern, pattern color=blue!40] (\x+\randx*0.5, \y+\randy*0.5) rectangle ++(0.5,0.5);
            \draw[black, thick, preaction={fill, blue!90}] (\x+\randx*0.5, \y+\randy*0.5) rectangle ++(0.5,0.5);
        }
    }

    \foreach \i in {0,0.5,...,4} {
        \draw[gray, thin] (\i,0) -- (\i,4); 
        \draw[gray, thin] (0,\i) -- (4,\i); 
    }

\end{tikzpicture}} &
        \resizebox{0.13\textwidth}{!}{\begin{tikzpicture}
    \pgfmathsetseed{72}

    \def\cells{4} 
    \def\size{4} 
    \def\smallsize{1} 

    \foreach \x in {0,1,...,\numexpr\cells-1} {
        \foreach \y in {0,1,...,\numexpr\cells-1} {
            \def\randx{1} 
            \def\randy{0}
            \def\pattern{north west lines} 

            \draw[black, thick, preaction={fill, red}, pattern=\pattern, pattern color=red!40] (\x,\y) rectangle ++(1,1);
            \draw[black, thick, preaction={fill, red!90}] (\x,\y) rectangle ++(1,1);

            \draw[black, thick, preaction={fill, blue}, pattern=\pattern, pattern color=green!40] (\x+\randx*0.5, \y+\randy*0.5) rectangle ++(0.5,0.5);
            \draw[black, thick, preaction={fill, blue!90}] (\x+\randx*0.5, \y+\randy*0.5) rectangle ++(0.5,0.5);
        }
    }

    \foreach \i in {0,0.5,...,4} {
        \draw[gray!50, thin] (\i,0) -- (\i,4); 
        \draw[gray!50, thin] (0,\i) -- (4,\i); 
    }

\end{tikzpicture}} \\
        \multicolumn{1}{c}{\scriptsize (a) Non-Regular Layout} &
        \multicolumn{1}{c}{\scriptsize (b) Regular Layout}
    \end{tabular}
    \caption{\footnotesize The proposed, sensor Layout in (a) with non-regular orientation of small pixel (green) and large pixel (red).
     In contrast, the regular layout in (b) with constant prototype pixel orientation.}
    \label{fig:layouts}
\end{figure}
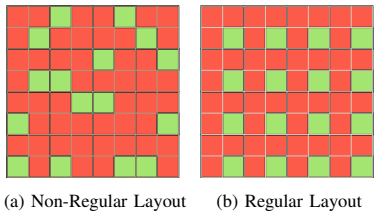

Inspired by Fujifilm's Super CCD \cite{SuperCCDSensor} this work also proposes a snapshot HDR sensor design with two prototype pixels of different sizes \cite{SuperCCDSensor}. 
However, unlike the Fuji sensor, the proposed sensor layout is able to acquire images aliasing-free and thereby maintain a high spatial resolution.
For this purpose non-regular sampling is suggested, which proofed to capture dominant scene frequencies, even with a reduced number of photometric measurements \cite{JSDE}. 
The considered sensor layout thereby combines the two imaging approaches and enables snapshot HDR imaging at a high spatial resolution.
To evaluate and demonstrate the impact of non-regular pixel arrangements on the resolution, the proposed snapshot HDR sensor, shown in.\ref{fig:layouts}a, is compared against a regularly arranged sensor with equivalent prototype pixels, illustrated in Fig.\ref{fig:layouts}b. 
\subsection{HDR Imaging with Spatially Varying Pixel Sizes}
\label{subsec:spatialVariation}
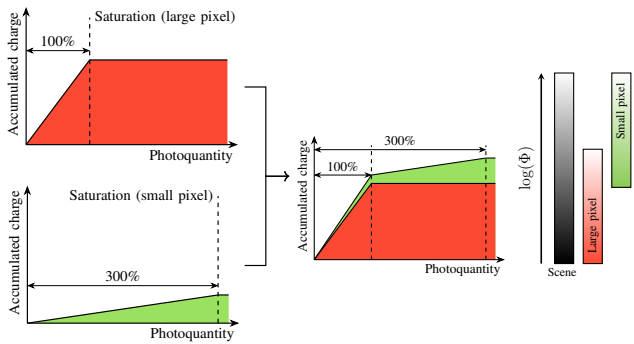
\begin{figure}[t]
    \centering
        \resizebox{0.48\textwidth}{!}{
\begin{tikzpicture}[node distance=3em]
    \node[scale=0.9] (c) {
        \begin{tikzpicture}[line join=round]
    \def\graphWidth{4.5cm}
    \def\graphHeight{3cm}
    \def\saturationPointA{1.5}
    \def\saturationPointB{4.5}
    \def\saturationChargeA{2}
    \def\saturationChargeB{0.666}

    \fill[blue] (0,0) -- (\saturationPointA,\saturationChargeA + 0.22) -- (\graphWidth,\saturationChargeB+\saturationChargeA) -- (\graphWidth + 0.25cm,\saturationChargeB+\saturationChargeA) -- (\graphWidth + 0.25cm,0) -- cycle;

    \fill[red] (0,0) -- (\saturationPointA,\saturationChargeA) -- (\graphWidth,\saturationChargeA) -- (\graphWidth + 0.25cm,\saturationChargeA) -- (\graphWidth + 0.25cm,0) -- cycle;

    \draw[thick, black] (0,0) -- (\saturationPointA,\saturationChargeA) -- (\graphWidth,\saturationChargeA)  -- (\graphWidth + 0.25cm, \saturationChargeA);

    \draw[thick, black] (0,0) -- (\saturationPointA,\saturationChargeA + 0.22) -- (\graphWidth,\saturationChargeB+\saturationChargeA) -- (\graphWidth + 0.25cm,\saturationChargeB+\saturationChargeA);

    \draw[dashed, thick] (\saturationPointA,0) -- (\saturationPointA,\graphHeight);
    \draw[dashed, thick] (\saturationPointB,0) -- (\saturationPointB,\graphHeight);

    \draw[<->, >=Stealth, black] (0,\saturationChargeA + 0.22) -- (\saturationPointA,\saturationChargeA + 0.22) node[above, midway]{\large 100\%};

    \draw[<->, >=Stealth, black] (0,\saturationChargeB + \saturationChargeA + 0.22) -- (\saturationPointB,\saturationChargeB + \saturationChargeA + 0.22) node[above, midway]{\large 300\%};

    \draw[->, >=Stealth, thick] (0,0) -- (\graphWidth + 0.5cm, 0) node[anchor=north east]{\large Photoquantity};
    \draw[->, >=Stealth, thick] (0,0) -- (0,\graphHeight + 0.25cm) node[rotate=90, anchor=south east] {\large Accumulated charge};

\end{tikzpicture}
        \begin{tikzpicture}
    \draw[-stealth, very thick] (-0.35,0) -- (-0.35,5) node[rotate=90] at (-0.75,2.5) {\Large $\log(\Phi)$}; 
    \shade[bottom color=black, top color=white] (0,0) rectangle (0.5,5);
    \draw[black, thick] (0,0) rectangle (0.5,5);

    \shade[bottom color=red, top color=white] (0.75,0) rectangle (1.25,3);
    \draw[black, thick] (0.75,0) rectangle (1.25,3);

    \shade[bottom color=blue!90!black, top color=white] (1.5,2) rectangle (2,5);
    \draw[black, thick] (1.5,2) rectangle (2,5);

    \node[below, align=center] at (0.2,0) {\normalsize Scene};
    \node[above, align=center, rotate=90] at (1.27,1.0) {\normalsize Large pixel};
    \node[below, align=center, rotate=90] at (1.5,4.0) {\normalsize Small pixel};
\end{tikzpicture}
    };
    \matrix[left=of c, row sep=0.2cm] (mat) {
        \node (a) {\begin{tikzpicture}
    \def\graphWidth{4.5cm}
    \def\graphHeight{3cm}
    \def\saturationPointA{1.5}
    \def\saturationPointB{4.5}
    \def\saturationChargeA{2}
    \def\saturationChargeB{0.666}

    \fill[red] (0,0) -- (\saturationPointA,\saturationChargeA) -- (\graphWidth,\saturationChargeA) -- (\graphWidth + 0.25cm,\saturationChargeA) -- (\graphWidth + 0.25cm,0) -- cycle;

    \draw[thick, black] (0,0) -- (\saturationPointA,\saturationChargeA) -- (\graphWidth,\saturationChargeA)  -- (\graphWidth + 0.25cm, \saturationChargeA);

    \draw[dashed, thick] (\saturationPointA,0) -- (\saturationPointA,\graphHeight) node[right]{\large Saturation  (large pixel)};

    \draw[<->, >=Stealth, black] (0,\saturationChargeA + 0.22) -- (\saturationPointA,\saturationChargeA + 0.22) node[above, midway]{\large 100\%};


    \draw[->, >=Stealth, thick] (0,0) -- (\graphWidth + 0.5cm, 0) node[anchor=north east]{\large Photoquantity};
    \draw[->, >=Stealth, thick] (0,0) -- (0,\graphHeight + 0.25cm) node[rotate=90, anchor=south east] {\large Accumulated charge};

\end{tikzpicture}};\\
        \node (b) {\begin{tikzpicture}
    \def\graphWidth{4.5cm}
    \def\graphHeight{3cm}
    \def\saturationPointA{1.5}
    \def\saturationPointB{4.5}
    \def\saturationChargeA{2}
    \def\saturationChargeB{0.666}

    \fill[blue] (0,0) -- (\graphWidth,\saturationChargeB) -- (\graphWidth + 0.25cm,\saturationChargeB) -- (\graphWidth + 0.25cm,0) -- cycle;

    \draw[thick, black] (0,0) -- (\graphWidth,\saturationChargeB) -- (\graphWidth + 0.25cm,\saturationChargeB);

    \draw[dashed, thick] (\saturationPointB,0) -- (\saturationPointB,\graphHeight) node[anchor=east]{\large Saturation (small pixel)};


    \draw[<->, >=Stealth, black] (0,\saturationChargeB + 0.22) -- (\saturationPointB,\saturationChargeB + 0.22) node[above, midway]{\large 300\%};

    \draw[->, >=Stealth, thick] (0,0) -- (\graphWidth + 0.5cm, 0) node[anchor=north east]{\large Photoquantity};
    \draw[->, >=Stealth, thick] (0,0) -- (0,\graphHeight + 0.25cm) node[rotate=90, anchor=south east] {\large Accumulated charge};

\end{tikzpicture}};\\
    };
    \draw[->, thick] (a.east) -- ++(0.5,0) |- (c.west);
    \draw[->, thick] (b.east) -- ++(0.5,0) |- (c.west);
\end{tikzpicture}
	}
\caption{\footnotesize Extended dynamic range of the proposed HDR sensor.}
\label{fig:combined_dynamic_range}
\end{figure}

The two prototype pixels interleave perfectly without any gaps or overlaps and achieve a 100\% fill factor of the image sensor, which is required for maximum light sensitivity.
Due to the two distinct sizes in light integration areas, a different amount of photons are collect by the two pixel types during one exposure.
Highlights and bright scene information are most accurately captured with the small pixel type.
While the large pixel with a threefold integration area size collects more photons during an exposure and is therefore capturing dark scene information that is otherwise dominated by noise.
It should be emphasised that the large light integration area is a physical mean to increase the light sensitivity and to lower the photometric limit of the image acquisition \cite{photometricLimit}. Thereby, the large pixel physically raises the signal strength by a factor of three whereas the energy of shot noise increases only with ${\sqrt 3}$ \cite{SNR}. This improves the overall SNR by ${\sqrt 3}$ and is very powerful in low-light scenarios.
As a drawback, the large pixel saturates earlier than the small pixel and highlights are likely to become blown out. The saturation behaviour is demonstrated in Fig. \ref{fig:combined_dynamic_range}, where the red saturation curve belongs to the large prototype pixel and the green curve shows the slower saturation behaviour of the small pixel types.
By combining the two differently-sized prototype pixels it is possible to benefit from the advantages of both pixel shapes and capture bright and dark areas simultaneously.
As indicated in Fig. \ref{fig:combined_dynamic_range}, the two prototype pixels cover different exposure ranges. Compared to a HR sensor, the large prototype pixel offers a pixel size ratio of 3:1 which yields an increase in dynamic range of 300\% for the proposed sensor, compared to the HR equivalent.

\subsection{Regular and Non-Regular Subsampling}
\label{subsec:Nonregular}
Non-regular sampling is known to be advantageous compared to regular sampling with respect to resolution per pixel \cite{NonRegularCMOS}. 
Due to the non-regularity, the aliasing becomes non-coherent and distributes like a noise floor over the spectrum \cite{Antialiasing}. 
Consequently, non-regular pixel placement is an established choice to enhance the spatial resolution with a limited number of photometric measurements \cite{JSDE}.
This work suggests to introduce non-regular placement of the two prototype pixels in the snapshot HDR sensor to circumvent aliasing artefacts in the image acquisition. 
The pixel placement defines how the sensor samples a scene thus the behaviour can be modelled by sampling a reference image. 
In this sense, the small prototype pixel is a point-wise measurement, while the large prototype pixel can be understood as an integrative measurement which bins three high resolution pixels.
The proposed non-regular sensor sampling pattern changes the orientation of the prototype pixels for each 2x2 HR pixel block, as indicated in Fig.\ref{fig:layouts}a.
Whereas the regular sampling approach assigns the small and the large prototype pixel always to a same fixed position within each 2x2 HR pixel block, as shown in Fig.\ref{fig:layouts}b.
Note, that the exact orientation within the 2x2 HR pixel block is secondary, as the major gains result from the non-regularity of the sampling.

Depending on the radiant flux of the scene, some photometric measurements will clip or will be dominated by noise. 
A well-exposed scene with medium scene intensities is correctly captured by both prototype pixels since the radiant flux is included in both fixed dynamic ranges. In this case both readings, from the small and the large pixel are considered as valid and the entire light-sensitive sensor plane gathers scene information.
 On contrary a low-light scene only produces a valid signal to noise ratio for the large pixel while the small pixel reading is indistinguishable from noise. In such lighting situations only 75\% of the sensor plane is providing information.
 For highlights the opposite holds true and only the small pixel reading is valid and the large pixel produces a saturated output. In such regions the local fill factor is reduced to only process information from 25\% of the photo sensitive area. This behaviour is summarised in Tab. \ref{tab:lighting}.
 As a consequence, of severe lighting conditions only a single prototype pixel reading is considered valid and the noisy or clipped value is discarded.
In the utmost case that the radiant flux of the scene exceeds the dynamic range of the camera sensor, both pixel types produce clipped or noise-dominated readings. In this case, the noisy value of the large pixel is kept during under-exposure, and for over-exposed scenes the saturated small pixel reading is considered for further processing.
 \begin{figure}[t]
 \vspace{-1mm}
	\centering
  	\includegraphics[width=0.48\textwidth]{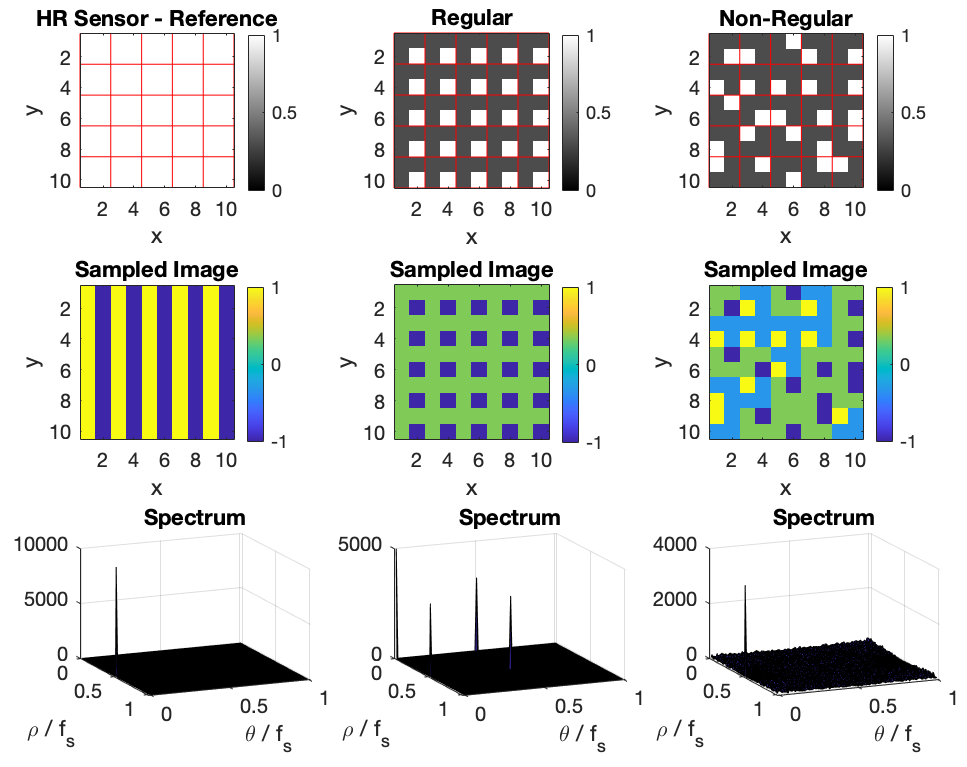}   
    \caption{\footnotesize Impact of pixel placement on the spatial resolution. 
    The top row shows a reference HR sensor (left), the regular HDR sensor (middle), and the proposed non-regular sampling pattern (right) which are used to sample a striped reference image. The middle row depicts the corresponding sampled image. Sampling with the HR sensor yield an output equivalent to the input reference. The respective Fourier spectra are shown in the bottom row.
    }
    \label{fig:regularity}
\end{figure}
\begin{table}[t]
\caption{ \footnotesize Relation between scene intensity and pixel readings.}
\centering
\begin{tabular}{|l|c|c||c|}
\hline
\textbf{Scene Intensity} & \textbf{Small Pixel} & \textbf{Large Pixel} & \textbf{Local Fill Factor} \\
\hline
High light & valid & discarded & 25\% \\
Mid intensity & valid & valid &  100\% \\
Low light & discarded & valid &  75\%\\
\hline
\end{tabular}
\label{tab:lighting}
\end{table}
In Fig. \ref{fig:regularity} a HR sensor layout is compared against a regular layout, and the proposed non-regular HDR sensor arrangement.
In the first row the sensor layouts are schematically depicted, where white pixels indicate the point-wise sampling of the small prototype pixel, which is equivalent to an HR pixel. The grey L-shaped pixels represent the integrative photometric measurement of the large prototype pixel.
A vertically striped reference image is sampled with all three sensors and the sampled image is shown in the second row for the respective patterns.
The sampling of an image signal $I(x,y)$ can be described as multiplication with a sampling pattern,
\begin{equation}
	\tilde{I}(x,y) = I(x,y) \cdot s(x,y)
\end{equation}
where the sampling pattern $s(x,y)$ is given by the pixel placement on the sensor plane.
For the HR sensor layout, the sampling frequency $f_s$ is large enough to capture the oscillation frequency of the stripes. The sampled image looks identical to the original and when analysing the spectrum a clear peak in $\rho$ direction appears in the Fourier domain.  
In case of the regular sampling it can already be seen in the spatial domain that the striped content is lost during the sampling process. This loss of information is confirmed in the Fourier domain where strong aliasing occurs which makes it impossible to recover the unique oscillation frequency of the original image.
In contrast, the non-regular pixel placement is able to capture the dominant spectral component since it changes the sampling distance over the image. As a result, the aliasing appears non-coherent and is scattered over the entire spectrum, contributing as a weak noise floor.
Similar results are obtained when changing the pixel orientation within the 2x2 HR pixel block. This demonstrates how a non-regular pixel arrangement can help to capture scene content aliasing-free even with integrative photometric measurements.  

For further processing and displaying a spatial varying deconvolution has to be applied in order to invert the integrative measurement of the large prototype pixel.
Likewise, for the sub-sampling cases where the scene intensities exceed the shared overlap in dynamic range, some pixels are discarded and scene information is missing and has to be reconstructed at these positions. 
Natural images have a sparse representations in the Fourier domain, and therefore  require only a few non-zero coefficients.
As a consequence, a reconstruction approach in the frequency domain such as the Joint Sparse Deconvolution and Extrapolation (JSDE) algorithm has proofed to be an effective choice \cite{JSDE} to deconvolve and extrapolate.
The JSDE builds a block-wise image model $g[m,n]$ in the Fourier domain
\begin{equation}
    g[m,n] = \sum_{k \in \mathcal{K}} c_k \varphi_k [m,n]
\end{equation}
where $m,n$ address the spatial coordinates within the block. This model approximates the measured image signal by iteratively adding Fourier basis functions $\varphi_k [m,n]$ to the model and weighting them individually according to an expansion coefficient $c_k$. 
The weighted residual energy is used as a criterion to evaluate the similarity between the model and the signal. A more detailed description of the reconstruction algorithm and the used, default parameters can be found in the original publication \cite{JSDE}.


\section{Simulation and Results}
\label{sec:results}
The image acquisition is simulated by sampling a HDR reference images with the different sensor sampling patterns. To model the saturation and noise thresholds, a simple camera model is introduced. Inspired by the CRF of a Canon 6D, the saturation threshold is set to 97\%  \cite{muliexposure}. The lower noise threshold is defined as 0.5\% of the upper clipping boundary \cite{datasetPaper}. The shot noise is implemented as multiplicative poisson noise while other noise sources are incorporated as additive gaussian noise.
  
\begin{figure}[t!]
\vspace{-1mm}
    \centering
    \setlength{\tabcolsep}{0.2em}
    \begin{tabular}{ccc}
        \includegraphics[width=0.155\textwidth]{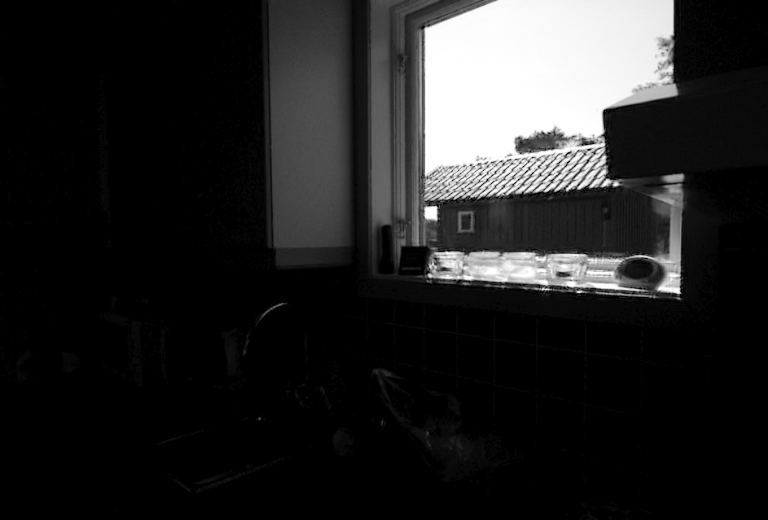} &
        \includegraphics[width=0.155\textwidth]{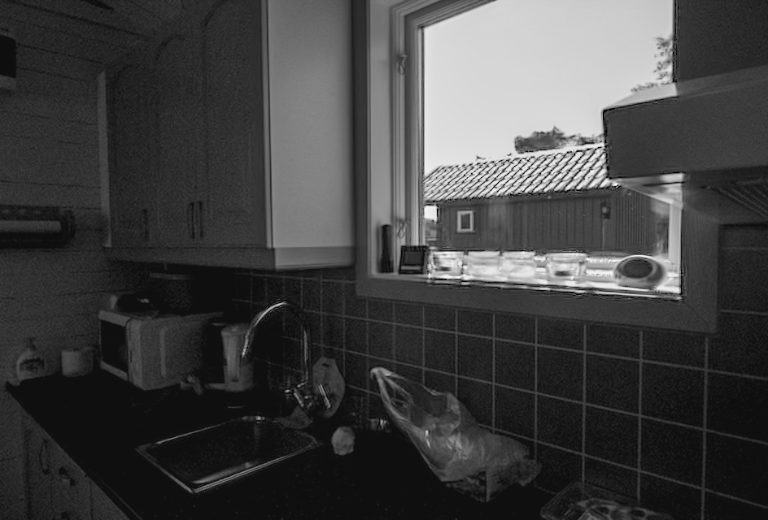} &
        \includegraphics[width=0.155\textwidth]{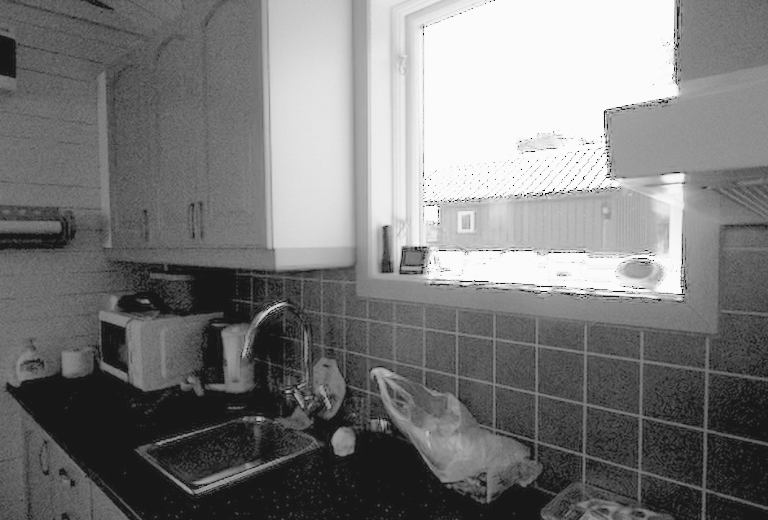} \\
        \includegraphics[width=0.155\textwidth]{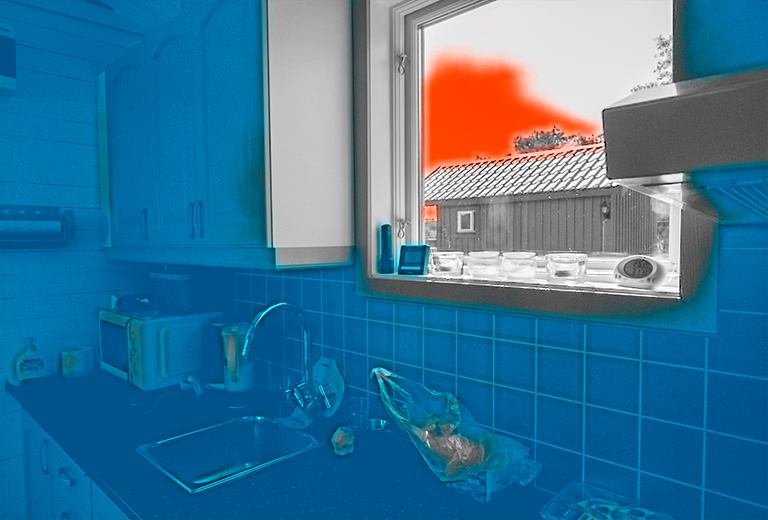} &
        \includegraphics[width=0.155\textwidth]{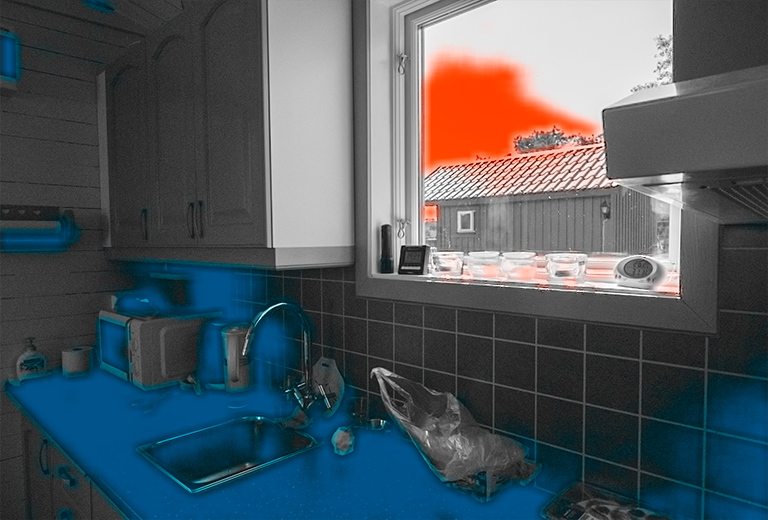}&
        \includegraphics[width=0.155\textwidth]{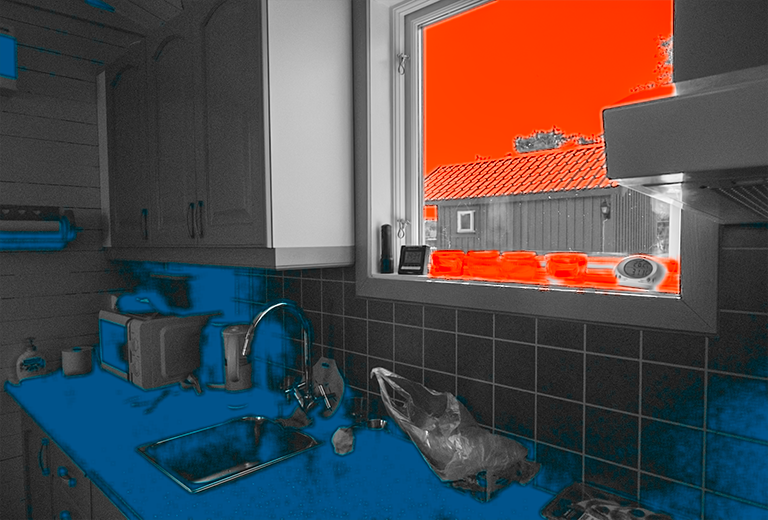}
         \\
		\noalign{\vspace{-0.3em}}
        \multicolumn{1}{c}{\scriptsize (a) Only small pixels} &
        \multicolumn{1}{c}{\scriptsize (b) Suggested HDR sensor } &
        \multicolumn{1}{c}{\scriptsize (c) Only large pixels} \\
    \end{tabular}
    \caption{\footnotesize Impact of saturation and noise thresholds.    
Readings from small pixels only (left), readings from large pixels only (right),  suggested HDR sensor combining both pixels (center). Blue: under-exposure; Red: over-exposure. Reinhard tone mapping is applied for visualisation \cite{reinhard}.}
   \label{fig:ShowThresholds}
\end{figure}

The effects of applying the camera model with the chosen thresholds to the sampling process are visualised in Fig. \ref{fig:ShowThresholds}, where the readings of the individual and combined prototype pixels are compared. 
The readings from the small pixels don't provide any information in the dark area while the large pixels yield clipped values in the bright scene parts. This is emphasised by the coloured image parts where blue indicates under exposure and red signals over exposure.
The HDR sensor which combines the readings from small and large prototype pixels yields a balanced representation of the scene where information is gathered in the dark and the bright areas. 
Depending on the measured intensities some pixel readings exceed the upper saturation threshold or fall below the lower noise threshold. These pixels readings are discarded during the sampling process. Consequently, a sparse image signal is obtained. To reconstruct the missing pixels at non-regular positions, and to deconvolve the binned, large prototype measurement, the Joint Sparse Deconvolution and Extrapolation (JSDE) algorithm is used \cite{JSDE}.
 Finally the reconstructed image quality is evaluated by comparing the reconstructed image against the original reference HDR image. 
 
Since the sensor layout evaluation is motivated from a signal processing perspective with particular emphasis on sustaining a high spatial resolution,  PSNR on the raw HDR image is used as an objective metric. In addition to that the PU21-PSNR is used which is converts the image into a uniform perspective domain before computing the PSNR \cite{PU21Metric}. Moreover, the HDR-VDP-3 probability detecting map (pmap) is used to provide insight about local quality changes by indicating where a human viewer is most likely to detect differences between the reconstructed image and its reference \cite{hdrvdp}.
\begin{figure}[t!]
    \centering
    \setlength{\tabcolsep}{0.15em}
    \begin{tabular}{cccc}
    \multirow{2}{*}[5.5em]{\rotatebox[origin=c]{90}{\small Non-Regular}} &
        \includegraphics[width=0.14\textwidth]{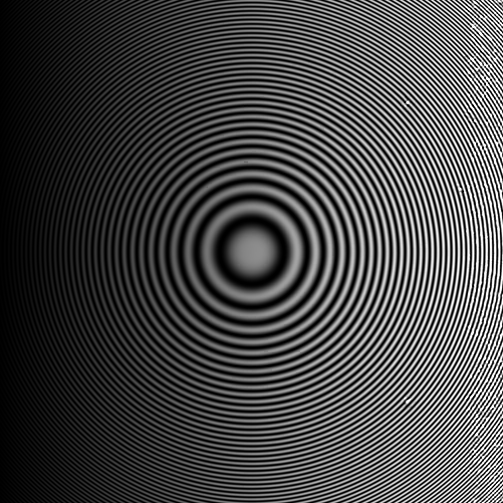} &
        
        \includegraphics[width=0.14\textwidth]{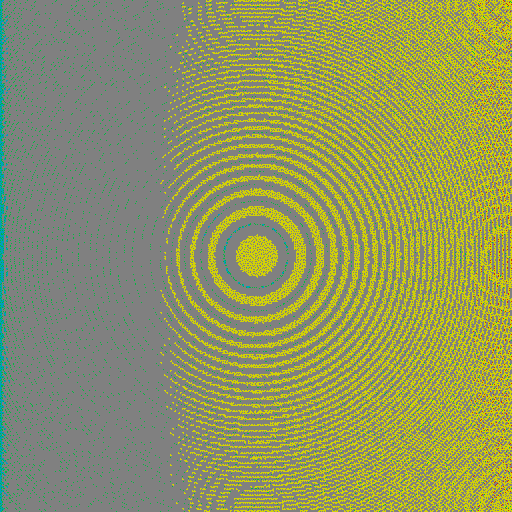} &
        \includegraphics[width=0.14\textwidth]{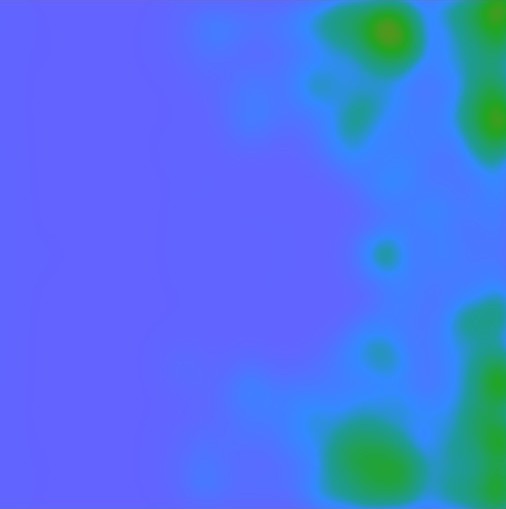} \\

        \multirow{2}{*}[5em]{\rotatebox[origin=c]{90}{\small Regular}}&
         \includegraphics[width=0.14\textwidth]{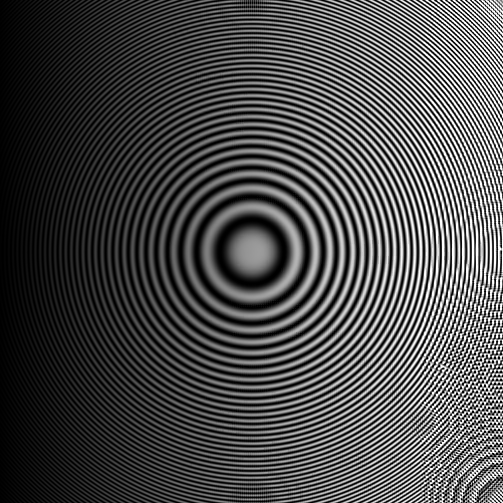} &
        \includegraphics[width=0.14\textwidth]{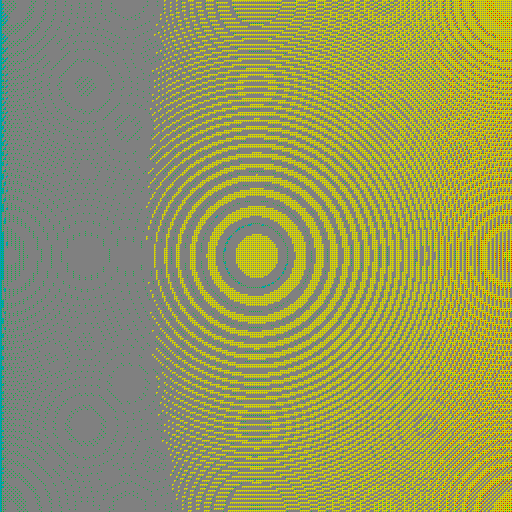}&
        \includegraphics[width=0.14\textwidth]{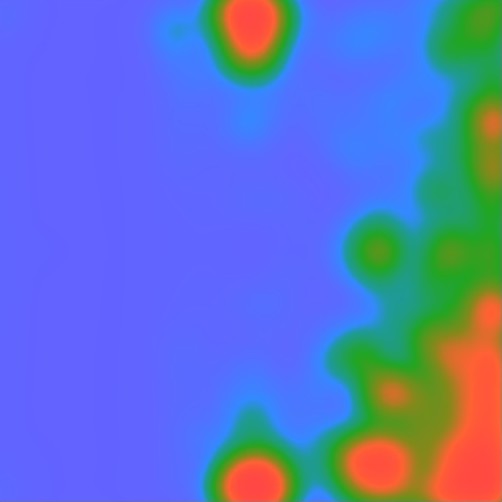} \\
		\noalign{\vspace{-0.3em}}

		&
        \multicolumn{1}{c}{\scriptsize (a) Sampled zoneplate} &
        \multicolumn{1}{c}{\scriptsize (b) Clipping mask} &
        \multicolumn{1}{c}{\scriptsize (c) HDR-VDP3 pmap} \\
    \end{tabular}
    \caption{\footnotesize Sampling a synthetic zoneplate with overlayed, horizontal gradient with the suggested HDR sensor in the regular and non-regular layout. 
    \textit{(Please pay attention, additional aliasing may be caused by printing or scaling. Best to be viewed enlarged on a monitor.)}}
   \label{fig:zoneplate}
\end{figure}
A synthetic zoneplate with an overlayed horizontal gradient has been generated to analyse the sensor's ability to capture high spatial frequencies and avoid aliasing artefacts.
The synthetic HDR image spans a luminance range from 0: $10^4$.
To compare the resolution achieved by a regular and non-regular sensor, this zoneplate is sampled with both sensor layouts.
When sampling the zoneplate with a regular pixel arrangement, the high-frequency parts of the rotation-symmetric chirp cannot be properly recovered.
 This is evident in Fig. \ref{fig:zoneplate}, where distortions and strong aliasing artefacts appear near the image borders, where the highest spatial frequencies are present. 
The clipping mask exhibits differently exposed areas with the colour code, red: small pixel over exposed, green: small pixel underexposed, yellow: large pixel overexposed, blue: large pixels under exposed.
Here the impact of the horizontal gradient can clearly be seen. The quality loss in bright areas is more drastic than in the very dark areas because, unlike most HDR methods, the proposed HDR sensor extends the dynamic range towards the lower end. Due to the included camera model, the brighter image part on the right hand side, causes more pixels to become saturated.
The discarded, clipped pixels reduce the sampling density in this area, which decreases the reconstructed image quality for both the regular and non-regular sensor pattern. This is confirmed by the HDR-VDP-3 probability detection map. Blue indicates very small probability of detecting a difference between the original and the reconstructed image. Green signals good quality, and red signals poor quality with a high probability of detecting differences.

When comparing the regular and the non-regular pmaps it can be seen that the non-regular sampling achieves a very good quality with a slightly higher probability to detect differences in the bright image region (c.f. green regions).
The quality map of the regular sampling follows a similar trend but here the HDR-VDP-3 metric identifies the aliasing artefacts as regions of poor quality where a human viewer is very likely to see differences. The asymmetric quality distribution in vertical direction of the regular pattern is a consequence of the fixed orientation of the prototype pixels.

In addition to the synthetic zoneplate, the suggested snapshot HDR sensor layout is evaluated on 50 natural 16-bit reference HDR images from the dataset of the Cambridge University \cite{datasetPaper}. The averaged image quality obtained with the regular and the suggested non-regular sensor pattern are listed in Table \ref{tab:results}, for the dataset and the zonplate.
Firstly, it should be mentioned that the images from the dataset contain less high frequency structures which explains the reduced difference between the regular and non-regular sensor arrangement for the dataset compared to the extreme case of the zoneplate.
When averaging the HDR-VDP-3 metric over the images from the dataset a score of 8.2\,JOD signals a very good quality for both compared HDR snapshot sensor layouts. Which indicates that that the two prototype pixels complement each other well and capture dark and bright scene parts jointly.
However, HDR-VDP-3 metric seems to be less sensitive to differences in the spatial resolution which is emphasised by the small quality difference of 0.5\,JOD for the zoneplate, while the non-regularly sampled PSNR scores a huge improvement of more than 5\:dB compared to the regularly sampled zoneplate.
When analysing the reconstructed image quality over the dataset, the PSNR metrics agree on a superiority of the non-regular pixel layout. The PSNR computed on the raw HDR data achieves an improved average of 0.32\:dB for the non-regular layout. The PU21-PSNR agrees with a superior quality for the non-regular sensor pattern and reads 0.6\:dB improvement in quality.
\begin{table}[t]
\centering
\caption{\footnotesize Average image quality of regular and non-regular HDR sensor.}
\begin{tabular}{|l|cc|cc|}
\hline
\multirow{2}{*}{\textbf{}} 
& \multicolumn{2}{c|}{\textbf{Si-HDR Dataset}\cite{datasetPaper}} 
& \multicolumn{2}{c|}{\textbf{HDR Zoneplate}}  \\
& \textbf{Regular} & \textbf{Non-Regular} 
& \textbf{Regular} & \textbf{Non-Regular}  \\
\hline
PSNR     & 40.43\:dB & 40.75\:dB & 40.07\:dB & 45.69\:dB  \\
PU21-PSNR    & 28.86\:dB & 29.50\:dB & 15.98\:dB & 19.01\:dB\\
HDR-VDP-3	 & 8.2\:JOD & 8.2\:JOD & 9.2\:JOD & 9.7\:JOD\\
\hline
\end{tabular}
\label{tab:results}
\end{table}
\section{Conclusion and Outlook}
\label{sec:conclusion}
In this paper a snapshot HDR sensor layout is presented which is based on two differently sized prototype pixels.
Prior HDR approaches that are based on spatially varying apertures offer increased dynamic range at the cost of introducing aliasing artefacts and thereby reducing the spatial resolution. This limitation is solved by arranging the pixels non-regularly in the proposed sensor layout. Due to the non-regularity aliasing artefacts are mitigated allowing to reconstruct very fine scene details. The image acquisition process with the proposed sensor layout is simulated and an extreme case scenario posed by a synthetic zoneplate is used to analyse the aliasing circumvention. When comparing the objective image quality, the non-regular HDR sensor outperforms its regular equivalent by more than 5\:dB in PSNR.
The results are backed up by an evaluation over 50 natural HDR images, suggesting the proposed HDR sensor as an effective layout for snapshot HDR imaging.
Now that the concept of the HDR snapshot sensor is established a following step will be to evaluate the sensor performance on colour imaging.
Moreover, the non-regular sensor could be extended to include a third, small prototype pixel with a neutral density filter to extend the cameras' dynamic range also towards the upper end.


\section*{Acknowledgment}

Funded by the Deutsche Forschungsgemeinschaft (DFG, German Research
Foundation) - Project number 516695992.

\pagebreak
\bibliographystyle{IEEEtran}
\bibliography{HDRImaging.bib}

@Book{HDRBook,
  title  = {High Dynamic Range Video – From Acquisition to Display and Applications},
  year   = {2016},
  author = {Dufaux, Frederic and Le Callet, Patrick and Mantiuk, Rafal and Mrak, Marta},
  month  = {04},
}

@InProceedings{Non-regularOptical,
  author  = {Schöberl, Michael and Belz, Alexander and Seiler, Jürgen and Foessel, Siegfried and Kaup, Andre},
  title   = {High dynamic range video by spatially non-regular optical filtering},
  year    = {2012},
  pages   = {2757-2760},
  month   = {09},
  journal = {Proceedings / ICIP ... International Conference on Image Processing},
}

@InProceedings{datasetPaper,
  author    = {Hanji, Param and Mantiuk, Rafa{\l} K. and Eilertsen, Gabriel and Hajisharif, Saghi and Unger, Jonas},
  title     = {Comparison of single image HDR reconstruction methods — the caveats of quality assessment},
  booktitle = {Special Interest Group on Computer Graphics and Interactive Techniques Conference Proceedings (SIGGRAPH '22 Conference Proceedings)},
  year      = {2022},
  doi       = {10.1145/3528233.3530729},
}

@Article{HDRCapture,
  author  = {Jacobs, Axel},
  title   = {High Dynamic Range Imaging and its Application in Building Research},
  journal = {Advances in Building Energy Research},
  year    = {2007},
  volume  = {1},
  pages   = {177-202},
  month   = {01},
}

@InProceedings{PU21Metric,
  author    = {Mantiuk, Rafa∤ K. and Azimi, Maryam},
  title     = {PU21: A novel perceptually uniform encoding for adapting existing quality metrics for HDR},
  booktitle = {2021 Picture Coding Symposium (PCS)},
  year      = {2021},
  pages     = {1-5},
  doi       = {10.1109/PCS50896.2021.9477471},
}

@Book{humanVision,
  title     = {High-dynamic-range (HDR) vision},
  publisher = {Springer},
  year      = {2007},
  author    = {Hoefflinger, Bernd},
}

@InProceedings{muliexposure,
  author    = {Debevec, Paul E. and Malik, Jitendra},
  title     = {Recovering high dynamic range radiance maps from photographs},
  booktitle = {Proceedings of the 24th Annual Conference on Computer Graphics and Interactive Techniques},
  year      = {1997},
  series    = {SIGGRAPH '97},
  pages     = {369–378},
  address   = {USA},
  publisher = {ACM Press/Addison-Wesley Publishing Co.},
  doi       = {10.1145/258734.258884},
  isbn      = {0897918967},
  numpages  = {10},
  url       = {https://doi.org/10.1145/258734.258884},
}

@InProceedings{Beamsplitter,
  author    = {Aggarwal, M. and Ahuja, N.},
  title     = {Split aperture imaging for high dynamic range},
  booktitle = {Proceedings Eighth IEEE International Conference on Computer Vision. ICCV 2001},
  year      = {2001},
  volume    = {2},
  pages     = {10-17 vol.2},
  doi       = {10.1109/ICCV.2001.937583},
}

@Book{DCG,
  title  = {High Dynamic Range Imaging: Sensors and Architectures},
  year   = {2013},
  author = {Darmont, Arnaud},
  month  = {01},
  isbn   = {9780819488312},
  doi    = {10.1117/3.903927},
  pages  = {1-124},
}

@InProceedings{SuperCCDSensor,
  author       = {Kazuhiko Takemura and Kazuya Oda and Toru Nishimura and Hiroshi Tamayama and Yutaka Takeuchi and Tetsuo Yamada},
  title        = {{Challenge for improving image quality of a digital still camera}},
  booktitle    = {Sensors and Camera Systems for Scientific, Industrial, and Digital Photography Applications IV},
  year         = {2003},
  volume       = {5017},
  pages        = {385 -- 392},
  organization = {International Society for Optics and Photonics},
  publisher    = {SPIE},
}

@InProceedings{NDFilters,
  author    = {Nayar, S.K. and Mitsunaga, T.},
  title     = {{High dynamic range imaging: spatially varying pixel exposures}},
  booktitle = {Proceedings IEEE Conference on Computer Vision and Pattern Recognition. CVPR 2000 (Cat. No.PR00662)},
  year      = {2000},
  volume    = {1},
  pages     = {472-479 vol.1},
  doi       = {10.1109/CVPR.2000.855857},
}

@Article{JSDE,
  author         = {Seiler, Jürgen and Jonscher, Markus and Ussmueller, Thomas and Kaup, André},
  title          = {{Increasing} {Imaging} {Resolution} by {Non}-{Regular} {Sampling} and {Joint} {Sparse} {Deconvolution} and {Extrapolation}},
  journal        = {IEEE Transactions on Circuits and Systems For Video Technology},
  year           = {2018},
  volume         = {29},
  pages          = {308-322},
  month          = {Jan},
  doi            = {10.1109/TCSVT.2018.2796725},
  faupublication = {yes},
  peerreviewed   = {Yes},
  url            = {https://arxiv.org/abs/2204.12867},
}

@Article{reinhard,
  author  = {Reinhard, Erik and Stark, Michael and Shirley, Peter and Ferwerda, James},
  title   = {Photographic Tone Reproduction For Digital Images},
  journal = {ACM Transactions on Graphics},
  year    = {2002},
  volume  = {21},
  month   = {05},
  doi     = {10.1145/566654.566575},
}

@InProceedings{SNR,
  author    = {Inochkin, Fedor and Kruglov, Sergey and Bronshtein, Igor},
  title     = {Increasing CCD frame rate and signal-to-noise ratio with high resolution capability using on-chip preprocessing and multisignal image representation},
  booktitle = {2016 IEEE NW Russia Young Researchers in Electrical and Electronic Engineering Conference (EIConRusNW)},
  year      = {2016},
  pages     = {209-213},
  doi       = {10.1109/EIConRusNW.2016.7448156},
}

@Article{photometricLimit,
  author  = {Schöberl, Michael and Brückner, Andreas and Foessel, Siegfried and Kaup, André},
  title   = {Photometric limits for digital camera systems},
  journal = {Journal of Electronic Imaging},
  year    = {2012},
  volume  = {21},
  pages   = {0501-},
  month   = {04},
  doi     = {10.1117/1.JEI.21.2.020501},
}

@Misc{hdrvdp,
  author        = {Rafal K. Mantiuk and Dounia Hammou and Param Hanji},
  title         = {HDR-VDP-3: A multi-metric for predicting image differences, quality and contrast distortions in high dynamic range and regular content},
  year          = {2023},
  archiveprefix = {arXiv},
  eprint        = {2304.13625},
  primaryclass  = {eess.IV},
}

@Article{Antialiasing,
  author  = {Mark A. Z. Dipp{\'e} and Erling Henry Wold},
  title   = {Antialiasing through stochastic sampling},
  journal = {Proceedings of the 12th annual conference on Computer graphics and interactive techniques},
  year    = {1985},
  url     = {https://api.semanticscholar.org/CorpusID:9774239},
}

@InProceedings{NonRegularCMOS,
  author    = {Yui Maeda and Junichi Akita},
  title     = {A CMOS image sensor with pseudorandom pixel placement for clear imaging},
  booktitle = {2009 International Symposium on Intelligent Signal Processing and Communication Systems (ISPACS)},
  year      = {2009},
  pages     = {367-370},
  doi       = {10.1109/ISPACS.2009.5383826},
}

\end{document}